\documentclass[preprint,aps,eqsecnum,superscriptaddress]{revtex4}
\usepackage[T1]{fontenc}
\usepackage[latin1]{inputenc}
\usepackage{graphicx,amsfonts}
\usepackage{amsmath}
\usepackage{slashed}

\newcommand{\be}{\begin{equation}}
\newcommand{\ee}{\end{equation}}
\newcommand{\bea}{\begin{eqnarray}}
\newcommand{\eea}{\end{eqnarray}}
\newcommand{\bml}{\begin{subequations}}
\newcommand{\eml}{\end{subequations}}
\newcommand{\pa}{\partial}

\newcommand{\vx}{\vec{x}}
\newcommand{\vy}{\vec{y}}

\newcommand{\vk}{\vec{k}}

\newcommand{\e}{\epsilon}

\newcommand{\ud}{{1\over 2}}
\newcommand{\Ket}[1]{\left|\, #1 \,\right\rangle }
\newcommand{\Bra}[1]{\left\langle #1 \right|}
\newcommand{\Bracket}[2]{\langle\, #1 \,|\, #2\,\rangle }
\newcommand{\rmi}{\mathrm{i}}

\begin{document}

\title {Noncommutative quantum mechanics: uniqueness of the functional description}

\author{F. S. Bemfica}
\author{H. O. Girotti}
\affiliation{Instituto de F\'{\i}sica, Universidade Federal do Rio Grande
do Sul, Caixa Postal 15051, 91501-970 - Porto Alegre, RS, Brazil}
\email{fbemfica, hgirotti@if.ufrgs.br}

\begin{abstract}
The generalized Weyl transform of index $\alpha$ is used to implement the time-slice definition of the phase space path integral yielding the Feynman kernel in the case of noncommutative quantum mechanics. As expected, this representation for the Feynman kernel is not unique but labeled by the real parameter $\alpha$. We succeed in proving that the $\alpha$-dependent contributions disappear at the limit where the time slice goes to zero. This proof of consistency turns out to be intricate because the Hamiltonian involves products of noncommuting operators originating from the non-commutativity. The antisymmetry of the matrix parameterizing the non-commutativity plays a key role in the cancelation mechanism of the $\alpha$-dependent terms.
\end{abstract}

\maketitle
\newpage

\section{Introduction}
\label{sec:level1}

In this work we shall be concerned with quantum systems whose dynamics is described by a self-adjoint Hamiltonian $H(Q,P)$ made up of the Cartesian coordinates $Q^j, j= 1,\ldots, N$ and their canonically conjugate momenta $P^j, j = 1,\ldots, N$. However, unlike the usual case, coordinates and momenta are supposed to obey the non-canonical equal-time commutation rules

\bml
\label{1}
\bea
&&\left[Q^l, Q^j\right] = -2 \rmi\hbar\theta^{lj}\,,\label{mlett:a1}\\
&&\left[Q^l, P^j\right] = \rmi\,\hbar\,\delta^{lj}\,,\label{mlett:b1}\\
&&\left[P^l, P^j\right] = 0\,.\label{mlett:c1}
\eea
\eml

\noindent
The distinctive feature is, of course, that the coordinate operators do not commute among themselves. The lack of non-commutativity of the coordinates is parameterized by the real antisymmetric $N \times N$ constant matrix $\|\theta\|$. In Refs.\cite{Chaichian1,Gamboa1,Gamboa2,Girotti1,Bemfica} one finds specific examples of noncommutative systems whose quantization has been carried out. The conditions for the existence of the Born series and unitarity were investigated in Ref.\cite{Bemfica1} while a general overview of the connection linking noncommutative theories with constrained systems was presented in \cite{Bemfica2}.

A realization of the algebra in Eq.(\ref{1}) can be obtained by writing

\bml
\label{II-11}
\bea
&& Q^l\,\equiv\,X^l\,+\,\theta^{lj}\,K^j\,,\label{mlett:aII-11}\\
&& P^l\,\equiv\,K^l\,,\label{mlett:bII-11}
\eea
\eml

\noindent
where the $X$'s and $K$'s obey the canonical commutation relations

\bml
\label{II-12}
\begin{eqnarray}
&&\left[X^{l}, X^{j}\right]\, = \,0\,,\label{mlett:aII-12}\\
&&\left[X^{l}, K^{j}\right] = \rmi\,\hbar\,\delta^{lj}\,,\label{mlett:bII-12}\\
&&\left[K^{l}, K^{j}\right] = 0\,,\label{mlett:cII-12}
\end{eqnarray}
\eml

\noindent
while repeated indices sum from $1$ to $N$. For a Hamiltonian

\be
\label{II-14}
H(P ,Q) = \frac{P^l P^l}{2 M} + V(Q)
\ee

\noindent
and, therefore,

\be
\label{II-15}
H(K^{l}\,,\,X^{l}\, + \,\theta^{lj}\, K^{j}) = \frac{K^l K^l}{2 M} + V(X^{l} + \theta^{lk}\, K^k)\,\equiv:\,H_{\theta}(K^{l}\,,\,X^{l})\,,
\ee

\noindent
it has been shown elsewhere\cite{Gamboa1,Gamboa2,Girotti1} that the time evolution of the system, in the Schr\"odinger picture, is described by the wave equation

\be
\label{II-16}
- \frac{\hbar^2}{2M} \nabla_{x}^2 \Psi(x,
t) + V (x)\star \Psi(x, t) = \rmi \hbar \frac{\pa
 \Psi(x, t)}{\pa t}\,,
\ee

\noindent
where $\nabla_{x}^2$ designates the $N$th-dimensional Laplacian, $M$ is a constant with dimensions of mass while $\star$ denotes the Gr\"onewold-Moyal product\cite{Gronewold,Moyal,Filk1}, namely,

\bea
\label{II-17}
V (x)\star \Psi(x, t)&\equiv& V(x) \left[\exp \left(- \rmi
\hbar \overleftarrow{\frac{\pa}{\pa
x^l}} \,\theta^{lj}\, \overrightarrow{\frac{\pa}{\pa
x^{j}}}\right)\right] \Psi(x, t)\nonumber\\
&=&\,V \left(x^{j}\,-\,\rmi \hbar \,\theta^{jl}\,\frac{\partial}{\partial x^{l}}\right)\Psi(x, t)\,.
\eea

The last section in Ref.\cite{Bemfica2} was concerned with the computation, in the case of the harmonic oscillator, of the Feynman kernel ${\mathcal K}(x_f , t_f ; x_{in} , t_{in} )$ by explicitly evaluating the corresponding phase space path integral. However, no effort was made there to determine whether this result is unique when the phase space integral is defined through the time slicing procedure \cite{Feynman,Tobocman,Davies,Garrod}. This is our main concern in this work.

In Section II we first pinpoint the main steps one goes through for implementing the time-slice definition of the phase space path integral yielding ${\mathcal K}(x_f , t_f ; x_{in} , t_{in} )$. In this regard, the generalized Weyl transform will be seen to play a central role. As it has long been recognized \cite{Cohen,Cohen1,Testa,Mayes}, ${\mathcal K}(x_f , t_f ; x_{in} , t_{in} )$ turns out not to possess a unique representation in terms of canonical path integrals. This lack of uniqueness becomes particularly critical for Hamiltonians involving products of non-commuting operators, which is an unavoidable feature of the Hamiltonian describing the quantum dynamics of a noncommutative system (see Eq.(\ref{II-15})). In Section III we demonstrate that, nevertheless, the antisymmetry of $\|\theta\|$ suffices for reestablishing uniqueness. Section IV contains the conclusions.

\section{The time-slicing procedure and the generalized Weyl transform}
\label{sec:level2}

We shall be looking for the path integral representation of the matrix element

\bea
\label{III-1}
{\mathcal K}_{\theta}(x_f,t_f ; x_{in},t_{in})=\langle{\vx_f}|\,e^{-\frac{\rmi}{\hbar}
(t_f-t_{in})H(P,Q)}|\,{\vx_{in}}\rangle
=\langle{\vx_f}|\,e^{-\frac{\rmi}{\hbar}
(t_f-t_{in})H_{\theta}(K,X)}\,|{\vx_{in}}\rangle\,,
\eea

\noindent
where Eqs.(\ref{II-14}) and (\ref{II-15}) have been taken into account.

The definition of the phase space path integral through the time slicing procedure demands that one starts by writing

\be
\label{III-2}
{\mathcal K}_{\theta}(x_f,t_f ; x_{in},t_{in})\,=\,\lim_{m\to\infty\atop\e\to0}\int dx_1\cdots dx_m\prod_{a=0}^m
{\mathcal K}_{\theta}(x_{a+1},t_{a+1} ; x_{a},t_a)\,,
\ee

\noindent
where $dx$ stands for $d^Nx$. The time interval $(t_f - t_{in})$ has been divided out into $m + 1$ subintervals of equal size. The limit $m \rightarrow \infty$ ($ \e \rightarrow 0$) must be understood as $\max (t_{a + 1} - t_a) \rightarrow 0$ while $\sum_{a = 0}^m (t_{a + 1} - t_a) = t_f - t_{in} \equiv: T$. Usually, ${\mathcal K}_{\theta}(x_{a+1},t_{a+1} ; x_{a},t_a)$ is referred to as the short-time propagator. Then, if in Cohen's general ordering scheme \cite{Cohen,Cohen2} one replaces $\hbar {\vec \theta} =  \vx$, $\hbar {\vec \tau} =  \vk$ and then sets

\be
\label{III-3}
f({\vec x} , {\vec k})\,=\,e^{\frac{\rmi}{\hbar}\alpha {\vec x}\cdot{\vec k}}\,,
\ee

\noindent
where $\alpha$ is a real dimensionless parameter such that

\be
\label{III-4}
-\,\frac{1}{2} \leq \alpha \leq +\,\frac{1}{2}\,,
\ee

\noindent
one arrives at \cite{Cohen2}

\bea
\label{III-5}
{\mathcal K}_\theta(x_{a+1},t_{a+1} ; x_a,t_{a})\,=\,(2\pi\hbar)^{-N}\int dk\,\exp \left\{ \frac{\rmi\e}{\hbar}\left[k_a^j{x^j_{a+1}-x^j_a\over\e}
-h_{\theta_{-\alpha}}\left(k_a,x_{a,a+1}(\alpha)\right)\right]\right\}\,.
\eea

\noindent
Here,

\be
\label{III-6}
x^j_{a,a+1}(\alpha)\,\equiv\,\left(\ud+\alpha\right)x^j_{a+1}+\left(\ud-\alpha\right)x^j_a
\,,
\ee

\noindent
$dk$ stands for $d^Nk$, whereas

\bea
\label{III-7}
&&h_{\theta_{-\alpha}}\left(k,x\right)\,\equiv\,(2\pi\hbar)^N {\mathrm tr}\left[H_{\theta}(K,X)
\Delta_{-\alpha}\left(\vec{K}-\vk , \vec{X}-\vx\right) \right]\nonumber\\
&&=\,(2\pi\hbar)^N\,\int dy\,\Bra{\vy}H_{\theta}(K,X)
\Delta_{-\alpha}\left(\vec{K}-\vk , \vec{X}-\vx\right)\Ket{\vy}\nonumber\\
&&=\,\int dy\,e^{\frac{i}{\hbar}{\vec k}\cdot{\vec y}}\,\Bra{\vx - \left(\frac{1}{2} + \alpha \right)\vy}\,H_{\theta}(K,X)\,\Ket{\vx + \left(\frac{1}{2} - \alpha \right)\vy}\,,
\eea

\noindent
is the generalized Weyl transform of index $\alpha$ (GWT$\alpha$) of the operator $H_{\theta}(K,X)$. It is a generalization of the Weyl correspondence rules \cite{Weyl,Mizrahi}. Furthermore,

\bea
\label{III-701}
&&\Delta_{\alpha}\left(\vec{K}-\vk , \vec{X}-\vx\right)\,\equiv\,(2\pi\hbar)^{-2N}\int\, d^N\gamma\, d^N\tau \,
e^{{\rmi \over \hbar}\left[\alpha\gamma^j\tau^j + \tau^j(K^j-k^j) +\gamma^j(X^j-x^j)\right]}\nonumber\\
&&=(2\pi\hbar)^{-N}\int \mathrm{d}\tau e^{-{i\over\hbar}\tau^i k^i}\Ket{\vx-\left(\ud+\alpha\right){\vec \tau}}
\Bra{\vx+\left(\ud-\alpha\right){\vec \tau}}\,.
\eea

\noindent
It is common use to symbolize the operation in Eq.~(\ref{III-7}) as

\be
\label{III-702}
H_\theta(K\,,\,X)\stackrel{-\alpha}{\longrightarrow}h_{\theta_{-\alpha}}\left(k,x\right)\,.
\ee

By putting everything together one arrives at

\bea
\label{III-8}
&&{\mathcal K}_{\theta}(x_f,t_f ; x_{in},t_{in})\,= \, \lim_{m\to\infty\atop\e\to0}(2\pi\hbar)^{-N(m+1)}\nonumber\\
&&\times \int \left(\prod_{a=1}^m dx_{a}\right)
\left(\prod_{a=0}^m dk_{a}\right)\,\exp \left\{ \frac{\rmi\e}{\hbar}\sum_{a=0}^m\left[
k_a^i\frac{x^i_{a+1}-x^i_a}{\e}\,-\,h_{\theta_{-\alpha}}\left(k_a,x_{a,a+1}(\alpha)\right)\right]\right\}\,,
\eea

\noindent
which defines the path integral through the time-slicing method. One is to notice that when inserting within the path integral the GWT$\alpha$ one must, correspondingly, evaluate the position dependent terms at the point $x^i_{a,a+1}(\alpha)$. Hence, the dynamics of a quantum mechanical system specified by a certain Hamiltonian operator does not have a unique translation into the path integral language. This is the converse of the operator ordering ambiguity arising when performing the quantization in according with the classical-quantum correspondence rules.

Of course, consistency requires that all the $\alpha$-dependence in the right hand side of Eq.(\ref{III-8}) should disappear after performing the limit $\e \rightarrow 0$. Up to our knowledge, no one has yet succeeded in carrying out such proof in full generality. In particular, and since the Hamiltonian $H_{\theta}$ involves products of noncommuting operators, the uniqueness of the path integral representation of ${\mathcal K}_{\theta} (x_f,t_f ; x_{in},t_{in})$ remains an open question for noncommutative systems. We shall address to this question in the next section.

\section{Functional formulation of the noncommutative quantum dynamics}
\label{sec:level3}

We start by looking for $h_{\theta_{- \alpha}}\left(k,x\right)$. The absence of ordering problems in the first monomial of the middle term in Eq.(\ref{II-15}) implies that

\be
\label{IV-1}
K^j K^j \stackrel{-\alpha}{\longrightarrow} \left(k^j k^j \right)_{-\alpha}\,=\,k^j k^j\,,
\ee

\noindent
as it can be checked by using Eq.(\ref{III-7}).

As for

\be
\label{IV-2}
V_{\theta}(K , X)\,\equiv\,V(X^{j} + \theta^{jl}\, K^l)
\ee

\noindent
the situation is far from being simple since it necessarily involves products of noncommuting operators. By starting from Eq.(\ref{III-7}) and after taking into account that

\bml
\label{IV-3}
\bea
\Bra{\vy}V\left(X^j+\theta^{jl}K^l\right)\Ket{\vk^{\,\prime}}&=&V(\vy)\star \Bracket{\vy}{\vk^{\,\prime}}\nonumber\\& =&
 V(\vy)e^{-\rmi\hbar \overleftarrow{\partial\over\partial y^j}\theta^{jl}\overrightarrow{\partial\over\partial y^l}}
\Bracket{\vy}{\vk^{\,\prime}}\,,\label{mlett:aIV-3}\\
\Bracket{\vy}{\vk}&\equiv&{1\over{(2\pi\hbar)^{N\over2}}}e^{{\rmi\over\hbar}y^jk^j}\,,\label{mlett:bIV-3}\\
\int d x\, \phi(\vx)\star\psi(\vx)&=&\int d x\, \phi(\vx)\,\psi(\vx)\,,\label{mlett:cIV-3}\\
\Bra{\vk^{\,\prime}}\Delta_{-\alpha}\left(\vec{K}-\vk , \vec{X}-\vx \right)\Ket{\vy}&=&
{(2\pi\hbar)^{{-3\over2}N}\over \left(-\alpha-\ud\right)^N}\nonumber\\
&\times& e^{{\rmi\over\hbar}\vy\cdot\left\{{-1\over\left(\alpha+\ud\right)}\vk-\left[1-{1\over \left(\alpha+\ud\right)}\right]\vk^{\,\prime}\right\}}
e^{{-\rmi\over\hbar\left(\alpha+\ud\right)}\vx\cdot(\vk^{\,\prime}-\vk)},\label{mlett:dIV-3}\\
{e^{{\rmi\over\hbar}k^{\prime j} y^j}\over (2\pi\hbar)^{N\over2}}
\star \Bra{\vk^{\prime}} \Delta_{-\alpha}\left(\vec{K}-\vk , \vec{X}-\vx\right)\Ket{\vy}&=&
{(2\pi\hbar)^{-2N}\over \left(-\alpha-\ud\right)^N}\,e^{{-\rmi\over \hbar\left(\alpha+\ud\right)}(\vx-\vy)\cdot(\vk^{\prime}-\vk)}\nonumber\\
&\times& e^{{-\rmi\over \hbar\left(\alpha+\ud\right)}\,k^{\prime j} \theta^{jl} k^l}\,,\label{mlett:eIV-3}
\eea
\eml

\noindent
one obtains

\bea
\label{IV-4}
&&V_{\theta}(K , X)\,\stackrel{-\alpha}{\longrightarrow} v_{{\theta}_{-\alpha}}(k , x)\,=\,(2\pi\hbar)^N\int\, dy\,\Bra{\vy}V\left(X^j+\theta^{jl}K^l\right)
\Delta_{-\alpha}\left(\vec{K}-\vk , \vec{X}-\vx\right)\Ket{\vy}\nonumber\\
&&=\,(2\pi\hbar)^N\int dy\, \int d k^{\prime}\, V(\vy)\left[{e^{{\rmi\over\hbar}k^iy^i}\over (2\pi\hbar)^{N\over2}}
\star \Bra{\vk^{\prime}}
\Delta_{-\alpha}\left(\vec{K}-\vk , \vec{X}-\vx\right)\Ket{\vy}\right]\nonumber\\
&&=\,{(2\pi\hbar)^{-N}\over \left(-\alpha-\ud\right)^N}\int dy\, V(\vy)\,e^{{\rmi\over \hbar\left(\alpha+\ud\right)}\,
\vk \cdot\, (\vx-\vy)}\,\int dk^{\prime} \,e^{{-\rmi\over \hbar\left(\alpha+\ud\right)}\,k^{\prime j}\,(x^j-y^j + \theta^{jl}k^l)}\nonumber\\
&&=\,V(x^j+\theta^{jl}k^l)\,e^{{-\rmi\over\hbar\left(\alpha+\ud\right)}k^j\theta^{jl}k^l}\,=\,V(x^j+\theta^{jl}k^l)\,.
\eea

This result is of paramount importance because it states that the would be $\alpha$-dependence of $v_{{\theta}_{-\alpha}}(k , x)$ is washed out by the antisymmetry of the matrix $\|\theta\|$. We emphasize that this would not be the case for an arbitrary $V(K , X)$, involving products of the noncommuting operators $K$ and $X$. Presently, where $K$ and $X$ enter into $V$ through the combination $X^j + \theta^{jl}K^l$, with $\theta^{jl} = - \theta^{lj}$, such dependence does not occur, this being true irrespective of the functional form of $V(x)$.

By returning with Eqs.(\ref{IV-4}) and (\ref{IV-1}) into (\ref{III-8}) one finds

\bea
\label{IV-5}
&&{\mathcal K}_{\theta}(x_{f},t_{f};x_{in},t_{in})\,=\,\lim_{m\to\infty\atop\e\to0}(2\pi\hbar)^{-N(m+1)}\nonumber\\
&&\times\,\int \left(\prod_{a=1}^m dx_{a}\right)
\left(\prod_{a=0}^m dk_{a}\right)\,e^{\frac{\rmi\e}{\hbar}\sum_{a=0}^m\left[
k^j_{a}\frac{x^j_{a+1}-x^j_a}{\e}-\frac{\vk^2_a}{2M}\,-\,V\left(x^j_{a,a+1}(\alpha)+\theta^{jl}k_{a}^{l}\right)\right]}\,,
\eea

\noindent
which still contains $\alpha$-dependent terms.

We are interested in proving the $\alpha$ independence of ${\mathcal K}_{\theta}(x_{f},t_{f};x_{in},t_{in})$ without imposing restrictions on the function $V(u)$, other than its analyticity at ${\vec u} = 0$. To this end, we start by introducing external sources for the coordinates and momenta, $J$ and $Z$, respectively. This enables us to rewrite Eq.(\ref{IV-5}) as

\be
\label{IV-6}
{\mathcal K}_\theta(x_{f},t_{f};x_{in},t_{in})
\, = \, \lim_{m\to\infty\atop \e\to0}\left.e^{-\frac{\rmi\e}{\hbar}\sum_{a=0}^m\,V({\vec L}_a)}\,
 W(J,Z,\e,m)\right|_{J=Z=0}\,,
\ee

\noindent
where

\be
\label{IV-7}
L^j_a \equiv \frac{\hbar}{\rmi\e}\left(\frac{\delta}{\delta J^j_{a}}+\theta^{jl}\frac{\delta}{\delta Z^l_a}\right)
\ee

\noindent
and

\bea
\label{IV-8}
&&W(J,Z,\e,m)\equiv (2\pi\hbar)^{-N(m+1)}\nonumber\\
&&\times\,\int \left(\prod_{a=1}^m dx_{a}\right)
\left(\prod_{a=0}^m dk_{a}\right)\,e^{\frac{\rmi\e}{\hbar}\sum_{a=0}^m\left[
k^j_{a}\frac{x^j_{a+1}-x^j_a}{\e}-\frac{\vk^2_a}{2M}+J^j_{a} x^j_{a,a+1}(\alpha)+Z^j_a k^j_{a}\right]}\,.
\eea

The momentum integrals in Eq.(\ref{IV-5}) are straightforward and after carrying them out one gets

\bea
\label{IV-9}
W(J,Z,\e,m)\,&=&\,(2\pi\hbar)^{-N(m+1)}\left({2M\pi\hbar\over \rmi\e}\right)^{{N\over2}(m+1)}\nonumber\\
&\times&\,\int \left(\prod_{a=1}^m dx_{a}\right)\,e^{\frac{\rmi M\e}{2\hbar}\left[A+{1\over\e}\sum_{a=1}^m\mu_a^ix_a^i
+{1\over\e^2}\sum_{a,b=1}^m x^i_a D_{ab}x_b^i\right]}\,,
\eea

\noindent
where

\bml
\label{IV-10}
\bea
A &\equiv& {2\over M}\left(\ud+\alpha\right)x^i_f J^i_m+{2\over M}\left(\ud-\alpha\right)x^i_{in} J^i_0
+\sum_{a=0}^m {\vec Z}^2_a\nonumber\\
&+&\,{2\over\e}x^i_f Z^i_m-{2\over \e} x^i_{in}Z^i_0\,+\,{\vx^2_f+\vx^2_{in}\over \e^2}\,,\label{mlett:aIV-10}\\
\mu_a^i & \equiv & -{2\over\e}x_{in}^i\delta_{a,1}-{2\over\e}x_{f}^i\delta_{a,m}+2\left(Z^i_{a-1}-Z^i_a\right)\nonumber\\
&+&\,{2\e\over M}\left[\ud\left(J^i_{a-1}+J^i_a\right)+\alpha\left(J^i_{a-1}-J^i_a\right)\right]\,,\quad a=1,\cdots,m\,,\label{mlett:bIV-10}\\
D_{ab} &\equiv& 2\delta_{ab}-\left(\delta_{a+1,\,b}+\delta_{a,\,b+1}\right)\,,\quad a,b=1\cdots, m\,.\label{mlett:cIV-10}
\eea
\eml

\noindent
The determinant and the inverse of the symmetric matrix $\|D\|$ can be computed at once and yield, respectively,

\bml
\label{IV-11}
\bea
&&\det(D_{ab})\,=\,m+1\,,\label{mlett:aIV-11}\\
&&D_{ab}^{-1}\,=\,{a(m-b+1)\over m+1}\,,\quad a\le b\,.\label{mlett:bIV-11}
\eea
\eml

\noindent
In turns, this enables us to perform the $x$-integrations in Eq.(\ref{IV-9}) with the result

\be
\label{IV-12}
W(J,Z,\e,m)=\left({M\over 2\pi \rmi\hbar T}\right)^{N\over2}\,e^{\Phi(J,K,\e,m)}\,
\ee

\noindent
where

\be
\label{IV-13}
\Phi(J,Z,\e,m)\equiv {\rmi\e\over \hbar}\left({MA\over2}-{M\over 8}\sum_{a,b=1}^m \mu^i_a D^{-1}_{ab}\mu^i_b\right)\,.
\ee

We now return with Eq.(\ref{IV-13}) into Eq.(\ref{IV-6}). After doing so, we shall be facing the problem of computing

\be
\label{IV-14}
\left[e^{-\frac{\rmi\e}{\hbar}\sum_{a=0}^m\,V({\vec L}_a)}\right]\,e^{\Phi(J,Z,\e,m)}\,=\,
\left[\sum_{r=0}^\infty{1\over r!}\left(-\frac{\rmi\e}{\hbar}\sum_{a=0}^m\,V({\vec L}_a)\right)^r\,\,\right]\,e^{\Phi(J,Z,\e,m)}\,.
\ee

\noindent
The analyticity of $V(u)$ at ${\vec u} = 0$ allows to write

\be
\label{IV-15}
V\left({\vec L}_a\right)\,=\,
\sum_{s=0}^\infty{1\over s!}\left.\frac{\partial^sV(u)}{\partial u^{i_1}\cdots\partial u^{i_s}}\right|_{u=0}
L^{i_1}_a\cdots L^{i_s}_a\,.
\ee

\noindent
Hence,

\be
\label{IV-151}
L^{i_1}_{a_1}L^{i_2}_{a_2}\cdots L^{i_v}_{a_v}\,e^\Phi
\ee

\noindent
is, up to coefficient functions not depending on the external sources, the form of a generic term entering the right hand side of Eq.(\ref{IV-14}). Now, since $\Phi$ is at the most bilinear in the external sources all monomials containing three or more factors $L$ applied to $\Phi$ vanish and, therefore, do not contribute to Eq.(\ref{IV-151}). To phrase it differently, only $L^i_a \Phi(J,Z,\e,m)|_{J=Z=0}$ and $L^i_a L^j_b \Phi(J,Z,\e,m)|_{J=Z=0}$ can survive  in Eq.(\ref{IV-151}). What remains to be done is to show that the just mentioned monomials are independent of $\alpha$.

To the above end, we start by recalling Eqs.(\ref{IV-10}), (\ref{IV-11}) and (\ref{IV-13}) and find

\bml
\label{IV-16}
\bea
&&{\hbar\over \rmi\e}{\delta\Phi\over\delta J_a^i}=
\delta_{a,m}\left[x^i_f+{\e\over T}\left(\ud-\alpha\right)\left(x^i_f-x^i_{in}\right)\right]
+\delta_{a,0}\left[x^i_{in}
+{\e\over T}\left(\ud+\alpha\right)\left(x^i_f-x^i_{in}\right)\right]\nonumber\\
&&\quad+\left(1-\delta_{a,0}-\delta_{a,m}\right)\left\{x^i_{in}\left(\ud+{m-2a\over m+1}\right)+
x^i_f{a\over m+1}+{\e\over T}\left[x^i_f+\alpha\left(x^i_f-x^i_{in}\right)\right]\right\}\nonumber\\
&&\quad-\e\sum_{b,c=1}^m\left[Z^i_{b-1}-Z^i_b+{\e\over M}\left(\ud+\alpha\right)J ^i_{b-1}
+{\e\over M}\left(\ud-\alpha\right)J ^i_{b}\right]D^{-1}_{bc}\nonumber\\
&&\quad \times\,\left[\left(\ud+\alpha\right)\delta_{c,a+1}+
\left(\ud-\alpha\right)\delta_{c,a}\right]\,,\label{mlett:aIV-16}\\
&&{\hbar\over \rmi\e}{\delta\Phi\over\delta Z_a^i}=
MZ^i_a+
{M\over T}\left(x^i_f-x^i_{in}\right)\nonumber\\
&&-M\sum_{b,c=1}^m\left[Z^i_{b-1}-Z^i_b+{\e\over M}\left(\ud+\alpha\right)J ^i_{b-1}
+{\e\over M}\left(\ud-\alpha\right)J ^i_{b}\right]
D^{-1}_{b,c}\nonumber\\
&&\times\,\left(\delta_{c,a+1}-\delta_{c,a}\right).\label{mlett:bIV-16}
\eea
\eml

\noindent
Therefore, in according with Eq.(\ref{IV-7}),

\bea
\label{IV-161}
&&L^i_a \Phi(J,Z,\e,m)|_{J=Z=0}=\delta_{a,m}x^i_f+\delta_{a,0}x^i_{in}+{M\over T}\theta^{ij}\left(x^j_f-x^j_{in}\right)\nonumber\\
&&+\left(1-\delta_{a,0}-\delta_{a,m}\right)\left[x^i_{in}\left(\ud+{m-2a\over m+1}\right)+
x^i_f{a\over m+1}\right]
+{\cal O}(\e)\,,
\eea

\noindent
where ${\mathcal O}(\e)$ embodies all terms vanishing when $\e\to0$. Thus, no $\alpha$-dependent terms survive in this limit.

Next on the line is $L^i_aL^j_b\Phi(J,Z,\e,m)$. We first notice that

\be
\label{IV-17}
L^i_aL^j_b\Phi(J,Z,\e,m)=\left({\hbar\over \rmi\e}\right)^2\left(
{\delta^2\over\delta J^i_a \delta J^j_b}+\theta^{ik}\theta^{jl}{\delta^2\over\delta Z^k_a \delta Z^l_b}
+\theta^{jl}{\delta\over\delta J^i_a}{\delta\over\delta Z^l_b}+
\theta^{il}{\delta\over\delta J^j_b}{\delta\over\delta Z^l_a}\right)\Phi
\ee

\noindent
and concentrate, afterwards, in computing each term in the right hand side of this last equation. We omit the details and quote

\bea
\label{IV-18}
&&\left({\hbar\over \rmi\e}\right)^2
{\delta^2\Phi\over\delta J^i_a \delta J^j_b}=
\delta^{ij}{\rmi\hbar\e\over M}\sum_{c,d=1}^m D^{-1}_{cd}
\left[{1\over 4}\left(\delta_{c,a+1}\delta_{d,b+1}+\delta_{a,c}\delta_{d,b+1}+\delta_{a+1,c}\delta_{d,b}+
\delta_{a,c}\delta_{b,d}\right)
\right.\nonumber\\
&&+\alpha\left(\delta_{a+1,c}\delta_{b+1,d}-\delta_{a,c}\delta_{b,d}\right)+\alpha^2\left(
\delta_{a+1,c}\delta_{b+1,d}-\delta_{a,c}\delta_{b+1,d}-\delta_{a+1,c}\delta_{b,d}+\delta_{ac}\delta_{bd}\right)\bigg]\nonumber\\
&&=\delta^{ij}{\rmi\hbar\e\over M}\begin{cases}
{m\over m+1}\left(\ud+\alpha\right)^2\,,\quad a=b=0\\
{m\over m+1}\left(\ud-\alpha\right)^2\,,\quad a=b=m\\
{\e\over T}\left\{{1\over 4}[4a(m-a)+m]+\alpha(m-2a)+m\alpha^2\right\}\,,\quad 0<a=b<m\\
{\e\over T}\left\{{1\over 4}[(4a+2)(m-b)+a+1]+\alpha(m-b-a)\right\}
\,,\quad 0<a<b<m\\
{\e\over T}\left\{{1\over 4}[(4b+2)(m-a)+b+1]+\alpha(m-b-a)\right\}
\,,\quad 0<b<a<m\\
\end{cases}\nonumber\\
&&=\delta^{ij}{\rmi\hbar\over 4MT}\begin{cases}
{\cal O}(\e)\,,\quad a=b=0\,,\,a=b=m\\
\e^2[4a(m-a)+m]+{\cal O}(\e)\,,\quad 0<a=b<m\\
\e^2[(4a+2)(m-b)]+{\cal O}(\e)\,,\quad 0<a<b<m\\
\e^2[(4b+2)(m-a)]+{\cal O}(\e)\,,\quad 0<b<a<m\\
\end{cases}\,,
\eea

\bea
\label{IV-19}
&&\left({\hbar\over \rmi\e}\right)^2\theta^{ik}\theta^{jl}{\delta^2\Phi\over\delta Z^k_a \delta Z^l_b}\nonumber\\
&&=\theta^{ik}\theta^{jk}{M\hbar\over \rmi\e}\sum_{c,d=1}^mD^{-1}_{cd}\left(\delta_{ac}\delta_{bd}-\delta_{a+1,c}\delta_{b+1,d}
+\delta_{a,c}\delta_{d,b+1}+\delta_{a+1,c}\delta_{b,d}-\delta_{ac}\delta_{bd}\right)\nonumber\\
&&=\theta^{ik}\theta^{jk}{M\hbar\over \rmi T}\,,\quad \forall\,a,b=0\cdots m\,,
\eea

\bea
\label{IV-20}
&&\left({\hbar\over \rmi\e}\right)^2\left(
\theta^{jl}{\delta\over\delta J^i_a}{\delta\over\delta Z^l_b}\right)\Phi\nonumber\\
&&=+{\theta^{ij}\hbar\over\rmi}\sum_{c,d=1}^m\left[
\left(\ud+\alpha\right)\delta_{a+1,c}
+\left(\ud-\alpha\right)\delta_{a,c}\right]
D^{-1}_{cd}\left(\delta_{d,b+1}-\delta_{d,b}\right)\,,
\eea

\noindent
and

\bea
\label{IV-21}
&&\left({\hbar\over \rmi\e}\right)^2\left(
\theta^{il}{\delta\over\delta J^j_b}{\delta\over\delta Z^l_a}\right)\Phi\nonumber\\
&&=-{\theta^{ij}\hbar\over\rmi}\sum_{c,d=1}^m\left[
\left(\ud+\alpha\right)\delta_{b+1,c}
+\left(\ud-\alpha\right)\delta_{b,c}\right]D^{-1}_{cd}
\left(\delta_{d,a+1}-\delta_{d,a}\right)\,.
\eea

Observe that a term like $\e^2[4a(m-a)+m]$, showing up in the right hand side of Eq.(\ref{IV-18}), is not eliminated by taking $\e\to0$. To exemplify why this happens we set, for instance, $a=m/2$ and find

\be
\label{IV-22}
\lim_{\e \to 0 \atop m \to \infty}\,\e^2\left[2 m\left(m-{m\over2}\right)+m\right]\, =\, T^2\,\neq \,0\,.
\ee

\noindent
Similarly, the second, third and fourth terms in the right hand side of Eq.(\ref{IV-18}) contain pieces that survive in the limit $\e \to 0$. However, all these contributions do not depend on $\alpha$. The same applies to the right hand side of Eq. (\ref{IV-19}), where the presence of the non-commutativity should be noticed. As for the right hand sides of Eqs. (\ref{IV-20}) and (\ref{IV-21}) they add up to

\bea
\label{IV-23}
\left({\hbar\over \rmi\e}\right)^2\left(\theta^{jl}{\delta\over\delta J^i_a}{\delta\over\delta Z^l_b}+
\theta^{il}{\delta\over\delta J^j_b}{\delta\over\delta Z^l_a}\right)\Phi
&=&\rmi\hbar\sum_{c,d=1}^mD^{-1}_{c,d}\left(\delta_{a+1,c}\delta_{b,d}-\delta_{a,c}\delta_{d,b+1}\right)\nonumber\\
&=&\begin{cases}
0\,,\quad a=b\\
\rmi\hbar\theta^{ij}{m+1+a-b\over m+1}\,,\quad 0\le a<b\le m\\
-\rmi\hbar\theta^{ij}{m+1+a-b\over m+1}\,,\quad 0\le b<a\le m\end{cases}\,,
\eea

\noindent
which are independent of $\alpha$. This completes the purported proof concerning the $\alpha$-independence of ${\mathcal K}_{\theta}(x_{f},t_{f};x_{in},t_{in})$.

It is worth mentioning that, individually, Eqs.~(\ref{IV-20}) and (\ref{IV-21}) contribute, among other things, $\alpha$-dependent terms. However, after these contributions are added up the just mentioned terms cancel out among themselves. The antisymmetry of $\| \theta\|$ is at the root of the cancelation mechanism.

\section{Conclusions}
\label{sec:level4}

The generalized Weyl transform of index $\alpha$ was successfully employed for implementing the time slice definition of the phase space path integral. As expected, this integral representation of the quantum dynamics turns out not to be unique but parameterized by the index $\alpha$. Of course, consistency demands all physical quantities being unique and, therefore, independent of $\alpha$.

We turned then into studying the Feynman kernel for the case of noncommutative quantum mechanics. The above mentioned lack of uniqueness appears to be particularly severe due to the fact that the Hamiltonian under analysis contains products of noncommuting operators. Unexpectedly, the antisymmetry of the matrix $\|\theta\|$, parameterizing the non-commutativity, saves the day. In fact, on the one hand, it kills the would be alpha dependence of the generalized Weyl transform while, on the other hand, it also takes care of the $\alpha$ dependence arising from the point on the slice where the coordinates are to be evaluated. The above holds true for all potentials $V(u)$ being analytic at ${\vec u} = 0$.

This work was partially supported by Conselho Nacional de Desenvolvimento Científico e Tecnológico (CNPq).

\end{document}